\documentclass[twocolumn,showpacs,preprintnumbers,amsmath,amssymb]{revtex4}
\usepackage{graphicx}% Include figure files
\usepackage{dcolumn}% Align table columns on decimal point
\usepackage{bm}% bold math
\begin{document}
\title{Optimal ranking in networks with community structure}
\author{Huafeng Xie$^{1,3}$, Koon-Kiu Yan$^{2,3}$, Sergei Maslov$^{3}$
\footnote{To whom the correspondence should be addressed: maslov@bnl.gov}}
\affiliation{
$^1$New Media Lab, The Graduate Center, CUNY
New York, NY 10016, USA\\
$^2$Department of Physics and Astronomy, Stony Brook University, \\
Stony Brook, New York, 11794, USA\\
$^3$Department of Physics, Brookhaven National Laboratory,
Upton, New York 11973,  USA}
%\altaffiliation{Department of Physics, Brookhaven National Laboratory,
%Upton, New York 11973,  USA}
%\author{Koon-Kiu Yan}
%\affiliation{Department of Physics, SUNY at Stony Brook, New York, 11794}
%\altaffiliation{Department of Physics, Brookhaven National Laboratory,
%Upton, New York 11973,  USA}
%\author{Sergei Maslov}
%\email{maslov@bnl.gov}
%\affiliation{Department of Physics, Brookhaven National Laboratory,
%Upton, New York 11973,  USA}
\date{\today}
\begin{abstract}
The World-Wide Web (WWW) is characterized by a strong community
structure in which groups of webpages (e.g. those devoted to a
common topic or belonging to the same organization) are densely
interconnected by hyperlinks. We study how such network
architecture affects the average Google rank of individual
communities. Using a mean-field approximation, we quantify how the
average Google rank of community webpages depends on
the degree to which it is isolated from the rest of the world in
both incoming and outgoing directions, and $\alpha$ -- the only
intrinsic parameter of Google's PageRank algorithm.
Based on this expression we introduce a concept of a web-community
being decoupled or conversely coupled to the rest of the network.
%
%web-community depending on whether its ranking is sensitive to the
%number of its connections with the rest of the WWW.
%
%numbers of hyperlinks connecting the
%community in
%question and the rest of the world in each of the two directions.
%
%In addition, we demonstrate that this dependence is sensitive to
%the value of $\alpha$., the only intrinsic parameter of Google's
%PageRank algorithm.
We proceed with empirical study of several internal
web-communities within two US universities. The predictions of our
mean-field treatment were qualitatively verified in those
real-life networks. Furthermore, the value $\alpha=0.15$ used by
Google seems to be optimized for the degree of isolation of
communities as they exist in the actual WWW.
\end{abstract}
 \maketitle

The World Wide Web (WWW) -- a very large ($\sim 10^{10}$ nodes)
network consisting of webpages connected by hyperlinks -- presents
a challenge for the efficient information retrieval and ranking.
Apart from the contents of webpages, the network topology around
them could be a rich source of information about their relative
importance and relevance to the search query. It is the effective
utilization of this topological information \cite{PageBrin} that
advanced the Google search engine to its present position of the
most popular tool on the WWW and a profitable company with a
current market capitalization around \$80 billion.
%%%%%%%%%%%%%%%%%%%%%%%%%
%To rank the
%importance of webpages, Google simulates the behavior of a large
%number of ``random surfers" who just follow a randomly selected
%hyperlink on each page they visit. The number of hits a given page
%gets in the course of such simulated process determines its
%ranking. It is intuitively clear that the larger is the number of
%hyperlinks pointing to a given webpage (its in-degree in the
%network), the higher are the chances of a random surfer to click
%on one of them and, therefore, the higher would be the resulting
%Google rank of this webpage. However, the algorithm goes beyond
%just ranking nodes based on their in-degrees. Indeed, the traffic
%directed to a given webpage along a particular incoming hyperlink
%is proportional to the popularity of the webpage containing this
%link. Therefore, the Google rank of a node is given by the
%weighted in-degree where the weight of each neighboring webpage
%reflects its importance and is determined self-consistently.
%%%%%%%%%%%%%%%%%%%
%The WWW is a very heterogeneous collection of webpages which can
As webpages can
be grouped based on their textual contents, language in which they
are written, the organizations to which they belong etc,
it should come as no surprise that the WWW has a strong
community structure \cite{community} in which similar pages are
more likely to contain hyperlinks to each other than to the
outside world. Formally a web community can be defined as a
collection of webpages characterized by an above-average
density of links connecting them to each other.%%%%%%%%%%%%

%%%%%%%%should be revised as this is no longer the key point
%In this letter we are going to address the question: how the
%community structure affects the Google rank of webpages inside the
%community. One might naively expect that the community structure
%always boosts the Google rank of its webpages as it tends to
%``trap'' the random surfer inside the community for a longer time.
%However, it turned out that it is not generally true. In fact the
%Google rank of community webpages could either increase or
%decrease with the density of inter-community links depending on
%the exact balance between average in- and out-degrees in the
%community. %%%%%%%%%%%%%%%%%%%%%%%%%%%%%%%%%%%%%%%%%%%%%%%%%%%%%
%%%%%%%%%%%%%%%%%%%%%%%%%%%%%%%%%%%%%%%%%%%%%%%%%%%%%%%%%%%%%%

In this letter, we are going to address the following question:
how does the relative isolation of community's webpages from the
rest of the network affects their Google rank? In addition we
would speculate the parameters of Google's PageRank algorithm were
selected for its optimal performance given the extent of the
community structure in the present WWW network.
%
%community structure affect the Google rank of webpages inside the
%community?
%
%We are going to present an analytical formula showing
%%%%%%%%%%%%%%%%%%%%%%%%%%should we say boost here????%%%%%%%%%%%%%
%that in principle the average Google rank of a community could be
%boosted up (or lowered down) by adjusting the number of links
%between the community and the rest of the world. More importantly,
%such a dependence of the average Google rank on network topology
%is sensitive to a parameter $\alpha$, which is the only intrinsic
%parameter in the PageRank algorithm.

In the heart of the Google search engine lies the PageRank
algorithm determining the global ``importance'' of every web page
based on the hyperlink structure of the WWW network around it.
When one enters a search keyword such as e.g. ``statistical
physics'' on the Google website the search engine first localizes
the subset of webpages containing this keyword and then simply
presents them in the descending order based on their PageRank
values. While the details of the PageRank algorithm have
undoubtedly changed since its introduction in 1997, the central
``random surfer'' idea first described in \cite{PageBrin} remained
essentially the same. From a statistical physics standpoint the
PageRank simulates an auxiliary diffusion process taking place on
the network in question. A large number of random walkers are
initially randomly distributed on the network and are allowed to
move along its directed links. Similar diffusion algorithms have
been recently applied to study citation and metabolic networks
\cite{Peterson} and the modularity of the Internet on the
hardware level represented by an undirected network of
interconnections between Autonomous Systems \cite{Maslov03}. As in
real web surfing, a random walker of the PageRank algorithm could
``get bored" from following a long chain of hyperlinks.
% and to decide to go to some other place in the web.
To model this
scenario, the authors introduced a finite probability $\alpha$ for
a random walker to directly jump to a randomly selected node in the network
not following any hyperlinks. This leaves the probability
$1-\alpha$ for it to randomly select and follow one of the
hyperlinks of the current webpage.
%The actual way to determine
%$\alpha$ is not revealed, but
According to \cite{PageRank}, in the real PageRank algorithm
$\alpha$ was chosen to be $0.15$. The algorithm then simulates
this diffusion process until it converges to a stationary
distribution. The Google rank (PageRank) $G(i)$ of a node $i$ is
proportional to the number of random walkers at this node in such
a steady state, and is usually normalized by
$\langle G(i)\rangle=1$. In this normalization, the flux of
walkers entering a given site due to random jump from all the
other nodes is given by $\sum_{i=1}^N \alpha G_i/N=\alpha$.
The continuity equation for this diffusion process reads
%determining the PageRank values $G(i)$ for all
%webpages in the WWW is thus
%\begin{equation}
$G(i)=\alpha+\sum_{j \to i} (1-\alpha) \frac{G(j)}{K_{out}(j)}$.
%\label{PR}
%\end{equation}
Here $K_{out}(j)$ denotes the number of hyperlinks (the
out-degree) of the node $j$ and the summation goes over all nodes
$j$ that have a hyperlink pointing to the node $i$. In the matrix
formalism the PageRank values are given by the components of the
principal eigenvector of an asymmetric positive matrix related to
the adjacency matrix of the network. Such eigenvector could be
easily found using a simple iterative algorithm. In order for this
one needs all nodes to satisfy $K_{out}(i)>0$. Practically, it is done by
iteratively removing pages with zero out-degrees from the network
\cite{PageRank}. Consider a network in which $N_c$ nodes form a
community characterized by an above-average density of edges linking
these nodes to each other. Let $E_{cw}$ to denote the total number of
hyperlinks pointing from nodes in the {\it community} to the
outside {\it world}, while $E_{wc}$ - the total number of
hyperlinks pointing in the opposite direction.
% (See Fig.\ref{cw_ill} for an illustration).
As the Google rank is
computed in the steady state of the diffusion process
%where the
%average number of random surfers currently visiting any given
%webpage does not change with time,
, the total current of surfers
$J_{cw}$ leaving the community must be
precisely balanced by the opposite current $J_{wc}$ of surfers
entering the community. Note that both $J_{cw}$ and
$J_{wc}$ consist of two contributions: the current via the direct
hyperlinks between the community and the outside
world, and the current due to random jumps.

Let $G_{c}=\langle G(i)
\rangle_{i \in C}$ to denote the average Google rank of webpages
inside the community. The average current flowing along a
hyperlink pointing away from the community
is given by $(1-\alpha) G_{c}/\langle
K_{out} \rangle_{c}$ and the total current leaving the community
along all those out-going links is $(1-\alpha) E_{cw}
G_{c}/\langle K_{out} \rangle_{c}$. The total number of random
walkers residing on nodes inside the community is $G_{c} N_{c}$
and the probability of a random jump to lead to a node outside the
community is $N_{w}/(N_{c}+N_{w})$, which is close to
$1$ as $N_{c}\ll N_{w}$. The contribution to the
outgoing current due to such jumps is given by $\alpha
G_{c}N_{c}$, and thus the total outgoing current is
$J_{cw} =(1-\alpha) G_{c} E_{cw} / \langle
K_{out}\rangle_{c} +\alpha G_{c} N_{c}$. Similarly the incoming
current $J_{wc}$ is given by $(1-\alpha)G_{w} E_{wc}/\langle
K_{out} \rangle_{w} +\alpha G_{w} N_{c}$. Equating these two
currents one gets $\displaystyle{\frac{G_{c}}{G_{w}}= \frac {
    (1-\alpha)E_{wc}/(\langle K_{out}\rangle_{w} N_c )
    +\alpha }
{
    (1-\alpha)E_{cw}/(\langle K_{out}\rangle_{c}N_c)
    +\alpha }.
}$ \noindent
One may notice that $\langle K_{out}\rangle_{w} N_c$
and $\langle K_{out}\rangle_{c}N_{c}$ are respectively equal to
$E_{wc}^{(r)}$ and $E_{cw}^{(r)}$ --
expected numbers of links connecting the community to the outside world
in a random network with the same degree sequence as the network in question
\cite{expect}.
%We denote these variables as $E_{wc}^{(r)}$ and
%$E_{cw}^{(r)}$ correspondingly.
By approximating $G_{w}\approx 1$,
we finally arrive at the following equation:
\begin{equation}
G_{c}=  \frac {
    (1-\alpha)\frac{E_{wc}}{E_{wc}^{(r)}}
    +\alpha }{(1-\alpha)\frac{E_{cw}}{E_{cw}^{(r)}}
    +\alpha }
    .
\label{eq2}
\end{equation}

For simplicity of notation, let us refer to the ratios
$E_{wc}/E_{wc}^{(r)}$ and $E_{cw}/E_{cw}^{(r)}$ as $R_{wc}$ and
$R_{cw}$ respectively. Roughly speaking, $R_{cw}$ and $R_{wc}$
quantify how isolated is a given community in both directions
connecting it to the outside world. In fact, in most communities
both ratios $R_{wc}$ and $R_{cw}$ are below $1$ because $E_{wc}$
and $E_{cw}$ are typically less than their expected values in a
randomized network \cite{Ecc}. One implication of the Eq.\ref{eq2}
is that the average Google ranking of a community depends on the
pattern of their connections with the outside world through the
ratios $R_{cw}$ and $R_{wc}$. For example if $R_{wc}$ is close to
$1$ (i.e. the number of links pointing to the community is roughly
the same as in a random network with the same degree
distribution), $G_c$ gets its maximum value $1/\alpha$ when
$R_{cw}\ll \alpha$, which could be interpreted as the community
very isolated in the out-direction. On the contrary, if the number
of out-going links from the community to the outside world is
roughly the same as in a corresponding randomized network, $G_c$
attains its minimum value of $\alpha$ if the community is very
isolated in the in-direction ($R_{wc}\ll\alpha$).
%Although the connections between the
%community and the outside world in both directions determine
%$G_c$, more importantly, it's
From Eq.\ref{eq2} one could easily see that the relative values of
isolation ratios $R_{cw}$, $R_{wc}$ and the parameter $\alpha$
determines the sensitivity of $G_c$ to community's connections with
the outside world. If either $R_{cw}$ or $R_{wc}$ is comparable to $\alpha$,
$G_c$ is sensitive to the exact number of links connecting the
community to the outside world in this particular direction.
Conversely, if both $R_{wc}, R_{cw}\ll \alpha$ the average Google rank of
community is no longer
sensitive to its outside connections, and its value is close to $1$ which is the
overall average value of $G_i$ for all nodes. In this case, we
would refer to this community as being ``decoupled" from the outside world. Of
course, whether a community is decoupled or coupled depends on the
value of $\alpha$. A community decoupled at a particular $\alpha$
could become coupled if a smaller $\alpha$ is chosen.
%%intra-community density
%increases imply R_cw R_wc decrease...this is a possbile location for this remark

\begin{figure}%reverse the alpha axis,%use the fitting curve?
[htbp] \centering
\includegraphics*[width=3.5in]{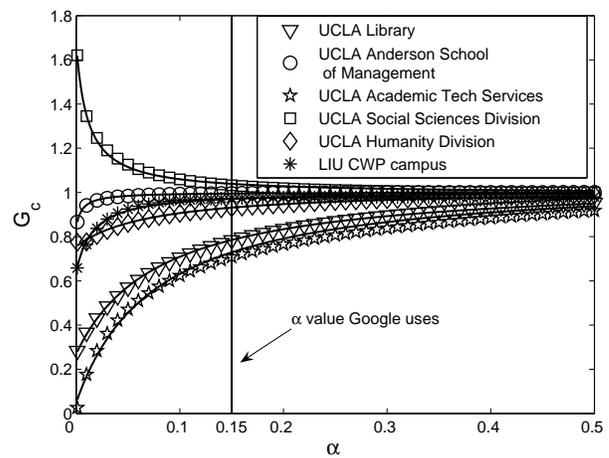}
\caption{The average Google rank $G_c$ of different communities as
a function of the parameter $\alpha$. The communities are within
real WWW networks of two US universities
(see Table \ref{table2} for details). The data points are obtained by running the PageRank
algorithm for different values of $\alpha$. Solid lines are two-parameter best
fits to the data with the Eq.\ref{eq2}.} \label{G_c_alpha}
\end{figure}

To empirically investigate the interplay between $G_c$ and $\alpha$ in real World-Wide
Web, we downloaded \cite{database} complete sets of hyperlinks contained in
all webpages within two US universities.
We then studied intra-university communities based either
on common interests (like schools or departments) or common
geographic locations (like individual campuses of a large university system).
(See Table \ref{table2} for details.) The relation between $G_c$ and $\alpha$ for six such
communities are shown in Fig.\ref{G_c_alpha}.
As expected from our calculations, as $\alpha$ is lowered
in all these communities $G_c$ starts to significantly deviate from $1$.
Moreover, the community ``UCLA social
science" deviates upward while all the others deviate downward.
This could be qualitatively explained by the Eq.\ref{eq2}, with the
observation that $R_{wc}$ is greater than $R_{cw}$ in this
community, while $R_{wc}$ is less than $R_{cw}$ in all the others
(see Table \ref{table}). Furthermore, by looking at which values
of $\alpha$ does $G_c$ starts to significantly deviate from $1$, one can see that
different communities become coupled to the outside
world for different $\alpha$'s. For example,
``UCLA Library" and ``UCLA Academic Tech. Service" reach the level
of $G_c=0.8$ when $\alpha$ is around $0.2-0.3$, while
``UCLA Anderson School of Management" and ``LIU CWP campus" reach
the same level of coupling only for much lower $\alpha\approx 0.01-0.05$.

We would like to point out that the Eq.\ref{eq2} is based on a
``mean-field" assumption. The average Google ranks and out-degrees
of community nodes sending links to the outside world are assumed
to be equal to the overall average values inside the community,
and the same is assumed for the nodes in the outside world that have
links to the community. Of course this is never perfectly true
for real web-communities. For example, a community may be linked from
the outside world by a highly ranked authority page, and receive
an in-coming current larger than predicted by our mean-field
calculation. Conversely, it can only get links from relatively
unimportant pages which would result in our mean-field model
overestimating the actual current. There is no
universal rule for estimating even the sign of the deviation from the mean
field predictions. Thus it is impossible to calculate
``corrections'' to our mean-field formula. Instead those corrections
have to be considered on a case-by-case basis. By allowing
parameters $R_{cw}$ and $R_{wc}$ in the
Eq.\ref{eq2} to deviate from their values prescribed by the mean-field theory
provides a simple mathematical formalism to quantify those corrections
for real communities. We define $R_{cw}^*$ and
$R_{wc}^*$ from the two-parameter best fit of the actual
$G_c(\alpha)$ dependence in a given community with the Eq.\ref{eq2} (see Table
\ref{table}.) One may regard $R_{cw}^*$ and $R_{wc}^*$ as
effective parameters, which in addition to simple geometrical
properties of the community such as numbers of links connecting it
to the outside world, take into account Google ranks of actual
pages sending those links. These ``renormalized'' ratios
$R_{cw}^*$ and $R_{wc}^*$ would be more accurate than their
``raw'' counterparts ($R_{cw}$ and $R_{wc}$) in determining
whether a particular web-community is coupled to or decoupled from
the outside world at a given value of $\alpha$.

%For the communities we have studied, the detail values are shown
%in Table \ref{table}.
\begin{table}
  \centering
  \caption{
  The basic statistics about the academic WWW networks downloaded
  from Ref. \cite{database}. We choose to study hyperlink networks
  within the Long Island University (LIU, 29476 nodes and 160457 edges) and separately within
  the University of California at Los Angeles (UCLA, 135533 nodes and 636595 edges).
  Following Google's original recipe \cite{PageBrin} we
  iteratively removed webpages with zero out-degree. The resulting networks
  consist of 15471 nodes and 90111 edges for the LIU and 31621 nodes and
  353370 edges for the UCLA. We then studied several large communities defined by
  the URL of their servers (e.g. .library.ucla.edu for the "UCLA Library"
  community.)}
%The topological parameters of communities are shown in the
%following table, where
%$N_c$ denotes the number of nodes in the
%community, $E_{cc}$ denotes the number of links inside the
%community and $E_{cc}^{(r)}$ denotes the same quantity in a
%randomized network with the same degree sequence.}
  \label{table2}
  \begin{tabular}{|c|c|c|c|c|c|}
  \hline
  Community & $N_c$& $E_{cc}$ & $E_{cc}^{(r)}$ & $E_{wc}$ & $E_{cw}$  \\
  \hline
  UCLA Library & 2028 & 23062 & 1699 &755 &2141  \\
  UCLA School of Management & 1340 & 15983 & 739 & 175 & 169 \\
  UCLA Academic Tech. Services & 1907 & 26597 & 2248 & 139 & 3113 \\
  UCLA Social Science Division& 626 & 3986 & 50 & 258 & 142 \\
  UCLA Humanity Division& 864 & 4846 & 79 & 397 & 445 \\
  LIU CWP Campus & 2756 & 18376 & 4105 & 336 & 1393 \\
  \hline
  \end{tabular}
\end{table}

\begin{table}
  \centering
  \caption{$R_{cw}$, $R_{wc}$, $R_{cw}^*$ and $R_{wc}^*$ for different
  communities. $R_{cw}$ and $R_{wc}$ are obtained by counting the
  links from the community to the world and vice versa, divided by
  the corresponding number of links in a random network with the
  same degree distribution \cite{expect}. $R_{cw}^*$ and $R_{wc}^*$
  are result of fitting the $G_c$ and $\alpha$ dependency
  via Eq.\ref{eq2}.}\label{table}
  \begin{tabular}{|c|c|c|c|c|}
  \hline
  Community & $R_{wc}$ & $R_{cw}$ & $R_{wc}^*$ & $R_{cw}^*$ \\
  \hline
  UCLA Library & 0.04 & 0.09 & 0.02 & 0.07 \\
  UCLA School of Management & 0.01 & 0.01 & 0.005 & 0.006 \\
  UCLA Academic Tech. Services & 0.007 & 0.1 & 0.003 & 0.07 \\
  UCLA Social Science Division& 0.04 & 0.03 & 0.02 & 0.01 \\
  UCLA Humanity Division& 0.04 & 0.08 & 0.05 & 0.07 \\
  LIU CWP Campus & 0.03 & 0.09 & 0.01 & 0.02 \\
  \hline
  \end{tabular}
\end{table}

\begin{figure}
[htbp] \centering
\includegraphics*[width=3.5in]{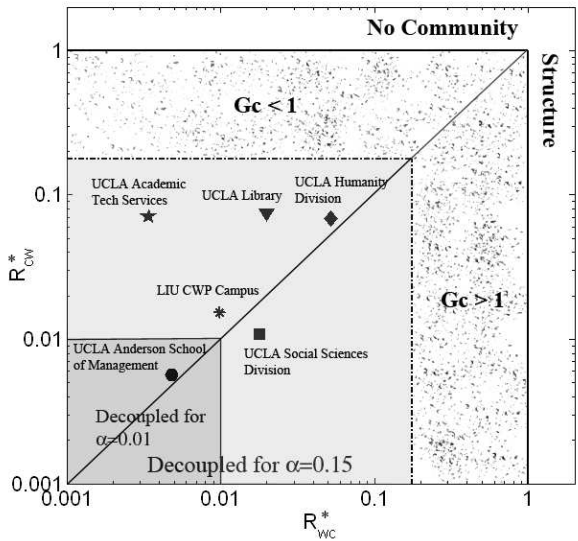}
\caption{$R_{cw}^*$ and $R_{wc}^*$ for different communities.
Communities inside the lightly shaded square are decoupled
from the rest of the world for $\alpha=0.15$, while the ones inside
the dark shaded square are decoupled for $\alpha=0.01$.}\label{2Dplot}
\end{figure}

The effective ratios $R_{cw}^*$ and $R_{wc}^*$ for the six
communities used in our study are listed in the Table \ref{table} and
visualized in Fig.\ref{2Dplot}.
Generally speaking, the closer to the origin is a community in
this figure, the lower is the value of $\alpha$ at which it first
becomes coupled to the outside world. One could see that for
$\alpha=0.15$, which is the actual value used by the Google
\cite{PageRank}, all of our six communities are essentially
decoupled from the outside world. However, if a much smaller value
of $\alpha$ (say $0.01$) is chosen, 5 out of 6 of our communities
(all except for the "UCLA Anderson School of Management") would
become sensitive to their connections with the outside world. In
principle, Fig.\ref{2Dplot} might be extended to include the region
where $R_{cw}^*$ and $R_{wc}^*$ are above one, but by definition
those points are not referring to well-defined communities. From
Eq.\ref{eq2} it follows that it is the asymmetry between $R_{cw}$ and
$R_{wc}$ which determines whether $G_c$ is greater than or
less than $1$. Thus the diagonal in Fig. \ref{2Dplot} separates
communities with $G_c> 1$ from those with $G_c < 1$. The ratio
between the $x$- and $y$-coordinates of the community in this plot
determines the asymptotic value of its Google rank $G_c$ for
$\alpha$ close to zero. Thus the two communities: ``UCLA Academic
Tech. Service" and ``UCLA Social Science", whose ratios between their
$x-$ and $y-$ coordinates in this plot are respectively the smallest and the largest
in our set deviate the most from $G_c=1$ as shown in
Fig.\ref{G_c_alpha}.

%In summary,  we employed a ``mean-field" approximation and derived
%analytically how the average Google rank of a community depends on
%the degree of its isolation from the outside world and the
%parameter $\alpha$.

%A possible practical consequence of our research might be that, a

The dominance of Google and the all-important role of its ranking
led to the appearance of services offering ``search engine
optimization'' to their clients. They promise to modify the
content and the hyperlink structure of client's webpages to
improve their Google rank. Our findings suggest one obvious way
how such an ``optimization'' could be achieved: the number of links
pointing to the outside world should be reduced to the minimum
while the number of intra-community
hyperlinks is kept at the maximum.
However, as we demonstrated above the success of such
a strategy depends on whether or not the community in question
is coupled to the outside world.
Indeed, the average Google rank of a decoupled community
is virtually insensitive to the exact balance of hyperlinks
connecting it to the outside world .
%
%the efficiency of such a search engine optimization technique
%depends on the value of $\alpha$, and it is only effective for
%those communities which are coupled to the rest of the world.
%
%Empirically we found that all of the intra-university communities
%used in our study are effectively decoupled for
%$\alpha=0.15$ but not for lower values of $\alpha$ (see Fig.\ref{2Dplot}).
%The existence of coupled
%communities and the resulting ability of webmasters to artificially
%boost the ranking of their communities
%is an undesirable effect for a search engine. Thus it should come
%as no surprise that the internal parameter $\alpha$ ultimately
%chosen by the Google's team would be highly optimized with respect
%to this effect.
%

Since coupling of web-communities to the outside world and the
resulting ability of their webmasters to artificially boost the
ranking is undesirable for a search engine, it should come as no
surprise that the internal parameter $\alpha$ chosen by the
Google's team is carefully selected to minimize this effect. To
make most of the communities decoupled the value of $\alpha$ in
the PageRank algorithm should be as large as possible. On the
other hand, for very large $\alpha$ the algorithm does not take
into account also the relevant network properties of the WWW.
Indeed for $\alpha$ close to 1, random surfers rarely follow
hyperlinks and thus nearly all topological information about the
network is lost. Therefore, the optimal value of $\alpha$ should
be chosen based on the realistic values of isolation parameters
$R_{cw}$ and $R_{wc}$. In our study we found all the communities
to be effectively decoupled at $\alpha=0.15$ but not at smaller
values of $\alpha$ (e.g $\alpha=0.01$ shown as a dark shaded
square in Fig.\ref{2Dplot}). Thus for our sample of
web-communities the value $\alpha=0.15$ proposed in
\cite{PageBrin} indeed optimizes Google's goals by striking the
best possible balance between the two opposing demands on the
value of $\alpha$.

Work at Brookhaven National Laboratory was carried
out under Contract No. DE-AC02-98CH10886, Division
of Material Science, U.S. Department of Energy.


\begin{thebibliography}{10}
\bibitem{PageBrin} S. Brin and L. Page,
%""The anatomy of a large-scale hypertextual {Web} search engine"
Computer Networks and ISDN Systems, {\bf 30}, 107 (1998).
%1--7,
%pages = "107--117",
%year = "1998",
%L. Page and S. Brin, in Proceedings of the Seventh International Web Conference
%(Elsevier, Amsterdam,1998).
%
%
\bibitem{community} R. Kumar, P. Raghavan, S. Rajagopalan,  and
A.Tomkins, Computer Networks {\bf 31}, 11 (1999).
%
%author = "Ravi Kumar and Prabhakar Raghavan and Sridhar
%Rajagopalan and Andrew Tomkins",
%    title = "Trawling the {Web} for emerging cyber-communities",
%    journal = "Computer Networks (Amsterdam, Netherlands: 1999)",
%    volume = "31",
%    number = "11--16",
%    pages = "1481--1493",
%    year = "1999",
\bibitem{Peterson}
%Topological properties of citation and metabolic networks
S. Bilke and C. Peterson Physical Review E {\bf 64}, 036106
(2001).
%
\bibitem{Maslov03} K. A. Eriksen, I. Simonsen, S. Maslov and K.
Sneppen, Phys. Rev. Lett. \textbf{90}, 148701 (2003).

\bibitem{PageRank} L. Page, S. Brin, R. Motwani and T. Winograd,
%"The PageRank Citation Ranking: Bringing Order to the Web"
Stanford Digital Library Technologies Project (1998).

\bibitem{expect} Indeed, in a random network
out of $\langle K_{out}\rangle_{w} N_w$
hyperlinks starting at nodes outside the community $\langle K_{out}\rangle_{w} N_w
N_c/(N_w+N_c) \simeq \langle K_{out}\rangle_{w} N_c$ would end up
pointing to community nodes. Similarly,
out of $\langle K_{out}\rangle_{c} N_c$
hyperlinks starting at community nodes $\langle K_{out}\rangle_{c}
N_c N_w/(N_w+N_c) \simeq \langle K_{out}\rangle_{c} N_c$ would
point to nodes in the outside world.

%As $N_c \ll N_w$, $\langle K_{out}\rangle_{c} N_c \simeq
%\langle K_{out}\rangle_{c} N_c
%\frac{N_w}{N_c+N_w}=E_{cw}^{(r)}$,\\
%$\langle K_{out}\rangle_{w}N_{c}\simeq \langle K_{out}\rangle_{w}
%N_c \frac{N_c}{N_c+N_w}=E_{wc}^{(r)}$

\bibitem{Ecc} Ususally communities
have higher than expected number of intra-community links:
$E_{cc}>E_{cc}^{(r)}$. Since $E_{cc}^{(r)}+E_{wc}^{(r)}=E_{cc}+E_{wc}
=N_c  \langle K_{in}\rangle_{c}$ and
$E_{cc}^{(r)}+E_{cw}^{(r)}=E_{cc}+E_{cw}=N_c  \langle K_{out}\rangle_{c}$,
this automatically implies that s
$E_{wc}<E_{wc}^{(r)}$ and $E_{cw}<E_{cw}^{(r)}$.
%=\sum{K_{in}(c)}$Starting from a random network, if links

\bibitem{database} Thelwall, M. Cybermetrics, Vol {\bf 6/7}, Issue 1. Paper 2 (2002-3).



%\bibitem{exact} Consider the subset of nodes $\mbox{in(C)}$ in the world
%whose nodes possess links pointing to the community C. For any
%node $i\in \mbox{in(C)}$, let $G(i)$ be the Google rank, and $f_i$
%be the fraction of links pointing to C. The sum $\sum_{i\in
%\mbox{in(C)}} G(i)f_i$ could be explicity evaluated. Under
%``mean-field" assumption, the sum is equvalent to $G_w N_c
%R_{wc}$. The numbers inside the brackets of Table \ref{table} in
%the column $R_{wc}$ are obtained by evaluting the sum for each
%community and divided by the corresponding $G_w N_c$. Similarly,
%by considering the subset of nodes in the community whose nodes
%possess links pointing to the world, we obtain the values
%corresponding to $R_{cw}$.

\end{thebibliography}
\end{document}